\documentclass[journal,draftcls,onecolumn,10pt]{IEEEtran}
\usepackage{graphicx} 

\usepackage[utf8]{inputenc}
\usepackage{amsfonts,amsmath}

\usepackage{subcaption}
\usepackage{epstopdf}
\usepackage{xcolor}

\usepackage[american]{babel}

\usepackage{etoolbox}

\definecolor{review}{RGB}{0,0,0}   

\definecolor{review_2}{RGB}{0,0,0}  
\definecolor{review_3}{RGB}{255,0,0} 

\DeclareMathOperator{\sinc}{sinc}

\begin{document}
%
\title {Combining AI/ML and PHY Layer Rule Based Inference - Some First Results}

%

\author{\IEEEauthorblockN{Brenda Vilas Boas$^{\star,\diamond}$, Wolfgang Zirwas$^{\star}$ and Martin Haardt$^{\diamond}$} \\
\IEEEauthorblockA{$^{\star}$Nokia, Munich Germany \\
Email: wolfgang.zirwas@nokia-bell-labs.com}\\
\and
\IEEEauthorblockA{
$^{\diamond}$Ilmenau University of Technology, Germany \\
Email: martin.haardt@tu-ilmenau.de}}

\maketitle

\begin{abstract}

In 3GPP New Radio (NR) 
Release 18 we see the first study item starting in May 2022, which will evaluate the potential of AI/ML methods for 
Radio Access Network (RAN)
1, i.e., for mobile radio PHY and MAC layer applications. 
We use the \textit{profiling} method for accurate iterative estimation of multipath component parameters for 
PHY layer reference, as it promises a large channel prediction horizon. We 
investigate options to partly or fully replace some functionalities of this rule based PHY layer method by AI/ML inferences, with the goal to achieve either a
higher performance, lower latency, or, reduced processing complexity. We provide first results for noise reduction, then a combined scheme for model order selection, compare options to infer multipath component start parameters, and, provide an outlook on a possible channel prediction framework.

\end{abstract}

%

\section{Introduction} 
Starting in May 2022, we see in 3GPP NR Release 18 the first study item \cite{AI_SI},  which will evaluate the potential of AI/ML methods for RAN 1, i.e., for mobile radio PHY and MAC layer applications. 
This study item "New SID on AI/ML for NR Air Interface" will assess the standard impact to limited use cases like channel prediction. For 6G we expect a fully integrated AI/ML air interface, which eventually might integrate all PHY building blocks into one deep transmitter and receiver~\cite{Hoydis}.

In the German funded project FunKI, we evaluate the potential benefits of AI/ML for the air interface in a somewhat broader sense. We want to understand, where and why it really makes sense to rely on ML- and/or PHY rule based methods. Even more, we are interested 
in identifying
a suitable framework, which combines the best of the two worlds. 
Similarly, as the 3GPP Release 18 study item, we selected as suitable use cases for our evaluations channel estimation and prediction and, as a side product, advanced positioning. 

There has been an extensive scientific effort on studying ML solutions for channel estimation and channel prediction. This is mainly due to the large range of possible neural networks (NNs) architecture configurations, such as  convolutional NNs and long-short term memory (LSTM) NN, 
as well as training procedures options, such as supervised learning and semi-supervised learning. For instance, authors in~\cite{20AhrensFading} propose a convolutional LSTM predictor where supervised learning is used to achieve a $10$ time steps ($5$~ms) prediction horizon for the entire frequency domain channel. In~\cite{21FischerRNN}, a combination of LSTM structures are proposed for channel estimation of time-frequency selective channels.
The work in~\cite{csinet} goes beyond channel estimation only and proposes a \textit{CsiNet} which jointly finds a compressed representation of the wireless channel and a decoder that can recover the original signal by using a variational auto-encoder (VAE) method.
Different to many other proposals, we do not try to directly infer the time domain channel evolution, but we aim
to estimate with high accuracy the underlying parameters of all $l~=~1 \ldots L $ multipath components (MPCs) like delay 
$ \tau_{l} $, amplitude $ \alpha_{l} $ and phase $ \varphi_{l} $. 
Based on at least two wireless channel observations, the channel prediction becomes then an extrapolation of these multipath component parameters into the future. 
Hence, with accurate MPC parameters estimation, there is a chance of reducing the CSI reporting overhead.

%
%
In this paper, we propose to combine ML and rule based PHY layer algorithms in a framework for an optimized \textit{profiling} implementation. 
The goal is to reduce complexity and processing time compared to the purely PHY based \textit{profiling} method~\cite{profiling}.
Our framework consists of a ML based noise reduction instance~\cite{22VilasUNN,21VilascGAN}, followed by a ML 
model order selection instance~\cite{21VilasMOS} with a pre-processing step from the higher order singular values decomposition (HOSVD)~\cite{08HaardtTensor} which is used in Unitary Tensor ESPRIT. Here, we focus on presenting the ML instance to estimate the MPCs' starting parameters and its impact on the \textit{profiling method}.
To the best of the authors' knowledge, this is the first work to combine ML and PHY functionalities specifically for parameter estimation task.
 
Below, Section~\ref{sec:framework} introduces the basic system setup and general integrated framework of PHY layer and ML blocks, Section~\ref{sec:noise_reduction} shortly discusses options for ML based noise reduction,     Section~\ref{sec:model_order} introduces a combined PH-ML scheme for improved model order selection,  Section~\ref{sec:start_parameters} 
presents the ML instance for estimating MPCs starting parameters,
while  Section~\ref{sec:channel_prediction} provides some results for the channel prediction use case. Section~\ref{sec:conclusion}  concludes the paper.

\section{Combined ML-PHY Framework} \
\label{sec:framework}
ML based channel prediction methods based on variational auto-encoders (VAE) like the already mentioned CsiNets \cite{csinet}, learn a transformation from CSI to a near-optimal number of representations (or codewords) and an inverse transformation from codewords to CSI. This raises the question:

\vspace{1.5mm}
\emph{Why AI/ML plus rule based PHY Inference?}
\vspace{1.5mm}

In the FunKI project we analyze alternative options for CSI reporting and channel prediction and are especially interested in possible combined PHY and AI/ML based inference of, e.g., multipath component parameters. We investigate the indirect channel prediction by means of multipath component parameters as these are reusable for multiple use cases, i.e., beside channel prediction also for accurate positioning, training of a digital twin, sensing, passive RADAR, etc. The combination of AI/ML with PHY layer rule based algorithms is motivated by the hope to get the best of two worlds:  

\begin{itemize}
	
	\item A deep knowledge of the PHY layer helps to understand potential gain mechanisms, steers ideas for advanced concepts, and allows to identify theoretical limits
	
	\item AI/ML promises to overcome rule based complexity issues, especially for iterative PHY layer algorithms
	
	\item Benefit from available, proven and efficient rule based methods like the Unitary Tensor ESPRIT algorithm 
	
	\item Generally, for the PHY layer there are many well understood methods, with predictable outcome, and limited or no need for training data
	
	\item Contrary, AI/ML promises smart and flexible control of optimization steps. In addition, it is always useful when training data are available, but deterministic algorithms are missing 
	
\end{itemize}

\begin{figure}[htbp]
	\centering
	\includegraphics[width=\columnwidth]{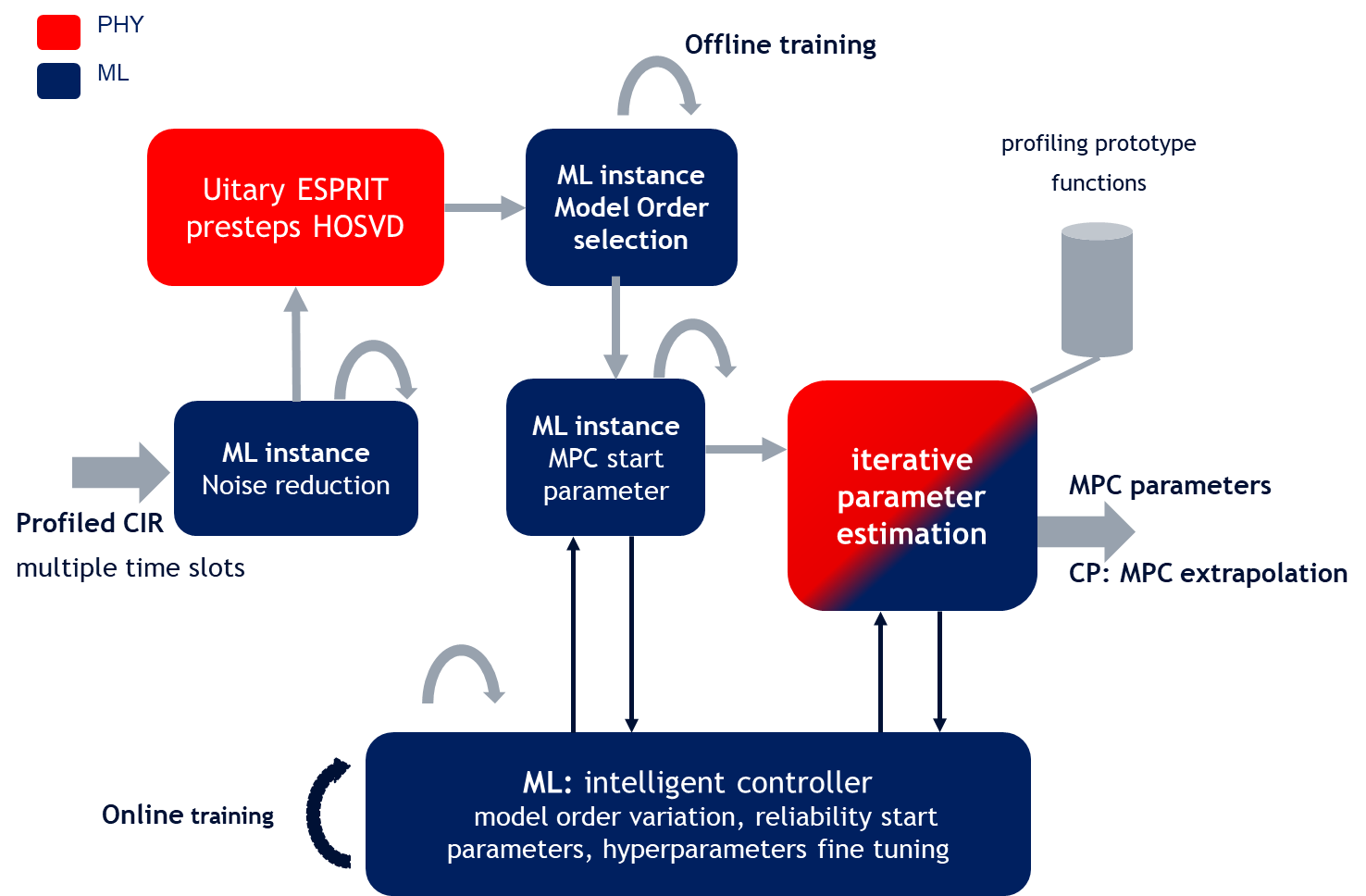}	
	\caption{\small{Illustration of the combined ML and PHY layer framework for channel prediction}}
	\label{fig:AI_FW}
\end{figure}

Based on this considerations, we defined a first combined AI/ML and PHY layer framework for channel prediction as illustrated in Figure~\ref{fig:AI_FW}%
, where red blocks correspond to rule based, or traditional, PHY layer algorithms, dark blue blocks refer to AI/ML algorithms and blocks with mixed red and dark blue colors represents a functionality that combines traditional and AI/ML algorithms. Some of the blocks operate in the frequency domain and others in the time domain, with the profiled CIR. This is clarified in the sections explaining each solution.

\subsection{System Model and basic Profiling Method for Channel Prediction} 
\label{sec:profiling}
For the detailed description of the original profiling method we refer to \cite{profiling}. Here, we provide only a high level illustration, which is derived for the evolution of a single radio channel component, similar as in \cite{ZTW+11}. We consider a massive MIMO OFDM system with $ N_{\mathrm{sc}}~=~600 $ subcarriers and a numerology similar to 3GPP LTE \cite{predictAntennas}. 
For the channel estimation we have one CSI RS per each of the $ M~=~50 $ physical resource blocks (PRB), i.e., one CSI RS every 180~kHz. The massive MIMO antenna is an array of overall $ N_{\mathrm{t}}~=~64 $ antenna elements forming a uniform rectangular array (URA) of size $~=~N_{1} \times N_{2}~=~4 \times 16 $ with a vertical spacing of $ d_{v}~=~0.7~\lambda $ and horizontal spacing of $ d_{h}~=~0.5~\lambda $. Here, $ \lambda $ is the wavelength of 0.14~m for the RF-frequency of 2.136~GHz. We represent the frequency domain channel transfer function as a tensor $ \mathbf{\mathcal{H}}~\in~\mathbb{C}^{N_{\mathrm{t}} \times M \times I } $, where $ I $ is the number of time instances taken into account. 
We can then define $ i_{\mathrm{v}}~=~1 \ldots N^{\mathrm{v}}_{\mathrm{beam}} $ vertical and $ i_{\mathrm{h}}~=~1 \ldots N^{\mathrm{h}}_{\mathrm{beam}} $ horizontal beam steering vectors:
 \begin{equation}
 \begin{array}{rrl} 
 \label{eq:V_b_weights}
 
	 \mathbf{v}^{\mathrm{v}}_{i_{\mathrm{v}}}(n_{1})	& = &    \mathrm{e}^{j 2 \pi (n_{1}-1)  \frac{d_{\mathrm{v}}}{\lambda} \sin [  \varphi_{\mathrm{v}}(i_{\mathrm{v}}) ] }; \ n_{1}~=~1 \ldots N_{1}; \\ 
	 
	 \mathbf{v}^{\mathrm{h}}_{i_{\mathrm{h}}}(n_{2})	& = &    \mathrm{e}^{j 2 \pi (n_{2}-1) \frac{d_{\mathrm{h}}}{\lambda}  \sin [ \varphi_{\mathrm{h}}(i_{\mathrm{h}}) ]  }; \ n_{2}~=~1 \ldots N_{2}; \\  

 \end{array} 
 \end{equation} 
 For $ \varphi_{\mathrm{v}} $ we used the set of tilt angles $ \{ 7^{\circ}, 12^{\circ} \} $ and for the horizontal beams the angles $ \{ -45^{\circ}, -15^{\circ}, 15^{\circ}, 45^{\circ} \} $ so that we regularly cover a $ 120^{\circ} $ cell sector.

Using the Kronecker product "$ \otimes $" we can apply the horizontal and vertical beams to the tensor $ \mathbf{\mathcal{H}} $ to all $ N_{\mathrm{sc}}~=~600 $ subcarriers and time instances $ I $ to get the wideband beamformed channel matrix:  
 \begin{equation}
 \begin{array}{rrl} 
 \label{eq:V_b_weights_per_beam}
 
	\mathbf{H}^{i_{\mathrm{v}},i_{\mathrm{h}}}_{m,i}	& = &  \mathbf{\mathcal{H}}^{T}(.,m,i) \ [ \mathbf{v}^{\mathrm{h}}_{i_{\mathrm{h}}} \otimes \mathbf{v}^{\mathrm{v}}_{i_{\mathrm{v}}} ];       \\

 \end{array} 
 \end{equation}
with $ \mathbf{H}^{i_{\mathrm{v}},i_{\mathrm{h}}}  \in \ \mathbb{C}^{ M \times I} $ as the beamformed frequency domain channel matrix. 
In this paper, we limit our evaluation to the case where, in a first step, the UE down-selects from all the beams, the one with the strongest average received power. 
As the selected beamformer is wideband, we can remove the superscripts and get $ \mathbf{H} $ for the beamformed frequency domain channel matrix.  

We ignore diffuse scattering and assume that the relevant part of the radio channel  $ \mathbf{H}(.,i) $ 
can be described by a deterministic set of multipath component parameters $ \Theta(t_{i}) $:
\begin{equation}
\begin{array}{rrl} 
\label{equ:parameters}

	\Theta(t_{i}) & = & \{ \tau_{1} \alpha_{1} \varphi_{1}, \ldots,   \tau_{l} \alpha_{l} \varphi_{l},  \ldots, \tau_{L} \alpha_{L} \varphi_{L}  \}  |_{t = t_{i}} ;\\

\end{array}
\end{equation}
The related over time $ t $ varying channel impulse response (CIR) is $ h_\mathrm{a}(\tau,t) $, which can be constructed for the time instances $ t_{i} $ from the parameter set $ \Theta(t_{i}) $ as:
\begin{equation}
\begin{array}{rrl} 
\label{equ:CIR}

	h_\mathrm{a}( \tau,t_{i})  &  =  &   \underset{l = 1 \ldots L}{\sum}   \alpha_l(t_{i}) \exp^{j \varphi_l(t_{i})}   \delta(\tau-\tau_l(t_{i}));  \\          
	\\
	\tau_l(t_{i})  &  =  &  \frac{d_l(t_{i})-d_0 }{c};  \\
	
\end{array}
\end{equation}
with $ c~=~3~10^{8} $~m/s as speed of light, $ d_l $ as the length of the multipath component $ l $, and $ d_0 $ as a reference distance for all multipath components. This discrete in time CIR $ h_\mathrm{a}(\tau,t_{i}) $ has to be filtered by the (single sided) bandwidth $ B $ of the RF-signal, which is in our case $ B $~=~10~MHz. 
Equivalently, we can apply in time domain a convolution with the sinc-function 
$\sinc( \tau' )~=~ \dfrac{\sin(\tau' B)}{\tau' B} $ with $ \tau'~\in~\mathbb{R} $: 
\begin{equation}
\begin{array}{rrl} 

	\bar{h}_\mathrm{a}(\tau,t_{i})  &  =  &   h_\mathrm{a}(\tau,t_{i})   \ast   \sinc(\tau');
\label{equ:CIR_convolution_1}

\end{array}
\end{equation} 
 The channel impulse response $ \bar{h}_\mathrm{a}(\tau,t_{i}) $ is sampled at the delay values of the time sample vector $ \mathbf{t}_{\mathrm{s}}~=~(\frac{1}{B} [1, \ldots  ,M N_{\mathrm{st}}]$/$N_{\mathrm{st}})^{T} $, where $ N_{\mathrm{st}} $ is the oversampling factor:
\begin{equation}
	\begin{array}{rrl} 
		\label{equ:CIR_convolution}

		\mathbf{\bar{h}}(t_{i})  &  =  &  | \mathbf{\bar{h}}_\mathrm{a}(\mathbf{t}_{\mathrm{s}},t_{i})  |;		
		
	\end{array}
\end{equation}
The $ | . | $ operator calculates the absolute value for the complex function values to get the profiled channel impulse response $ \mathbf{\bar{h}}~\in~\mathbb{R}_{+}^{M N_{\mathrm{st}} \times I} $, which will be then the input to our parameter estimators. 

The goal is to estimate the optimum parameter set $ \hat{\Theta}(t_{i}) $, which recreates $ \mathbf{\bar{h}}(t_{i}) $ as accurate as possible. 
For that purpose, we define the oversampled discrete delay vector $ \mathbf{t}'_{\mathrm{s}}~=~(\frac{1}{B} [1, \ldots, 2 M N_{\mathrm{st}}]  $/$ N_{\mathrm{st}})^{T} $ and replace Equation~(\ref{equ:CIR_convolution}) and Equation~(\ref{equ:CIR}) by its discrete implementation to get $ \mathbf{\bar{h}}^{'} $:
\begin{equation}
\begin{array}{rrl} 
\label{equ:CIR_convolution_discrete}

	\mathbf{\bar{h}^{'}}(t_{i})  &  =  &  | \underset{l = 1 \ldots L}{\sum}   \mathcal{Q}(\alpha_l) \exp^{j \mathcal{Q}(\varphi_l)}  \:   \sinc( \mathbf{t}'_{\mathrm{s}} - \mathcal{Q}(\tau_l) ) |; 

\end{array}
\end{equation} 
The function $ \mathcal{Q}(.) $ is then a suitably adapted quantizer applied to the multipath component parameters of $ \hat{\Theta}(t_{i}) $ and we replaced the convolution operation "$ \ast $" by a multiplication as the sync-function is symmetric.

One possible cost function for the optimizer is then to minimize the mean error signal power $ \varepsilon $ between the profiled channel impulses $ \mathbf{\bar{h}}_\mathrm{a}(t_{i},\tau_{s}) $ as the ground truth and the artificially constructed $ \mathbf{\hat{\bar{h}}}(t_{i}) $ based on the inferred or estimated set of multipath component parameters $ \hat{\Theta}(t_{i}) $:
 \begin{equation}
 \begin{array}{rrl} 
 \label{equ:CIR_error_signal} 

 \varepsilon  &  =  &   \underset{\tau_{s} = w_{\mathrm{start}} \ldots w_{\mathrm{stop}}}{\sum} | \mathbf{\bar{h}}_\mathrm{a}(\tau_{s},t_{i}) - \mathbf{\hat{\bar{h}}}(\tau_{s},t_{i})  |^{2};

 \end{array}
 \end{equation} 
The parameters $ w_{\mathrm{start}} $ and  $ w_{\mathrm{stop}} $ define then an evaluation window for the parameter inference. An alternative cost function to be minimized is the mean square error between the estimated parameter set $ \hat{\Theta}(t_{i}) $ versus the ground truth set $ \Theta(t_{i}) $. Note this ground truth set $ \Theta(t_{i}) $ is, for example, known in case of artificially generated radio channels for the purpose of training and/or evaluation.

For channel prediction, we use the estimated parameter sets $ \hat{\Theta}(t_{i}) $ for time instances $ t_{i}~=~t_{1} \ldots t_{\mathrm{ob}} $ and apply for each parameter a spline interpolation as well as extrapolation for the time instances $ t_{i}~>~t_{\mathrm{ob}} $. To get the predicted CSI, we can directly insert the predicted parameter sets $ \hat{\Theta}(t_{i})|_{t_{i}>t_{\mathrm{ob}}} $ into Equation~(\ref{equ:CIR_convolution_discrete}).

\section{Noise Reduction} 
\begin{figure}[tb!]
    \centering
    \includegraphics[width=0.7\columnwidth]{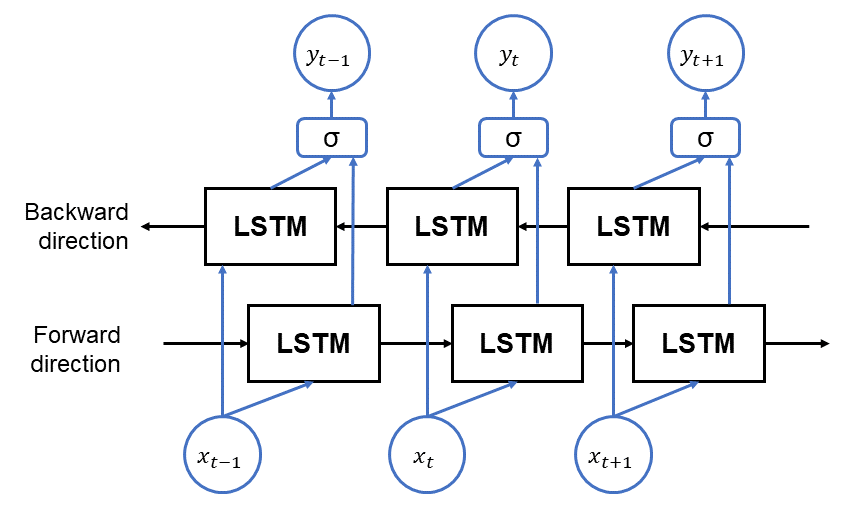}
    \caption{\small{A bidirectional LSTM layer combine LSTM processing units in the forward and backward directions. Here, we show the interconnection of $3$ bidirectional LSTM units with $\mathbf{x}$ as input vector and $\mathbf{y}$ as output.}}
    \label{fig:bilstm}
\end{figure}

\
\label{sec:noise_reduction}
The first block of our framework is the noise reduction block. Here, we choose to perform this task using AI/ML algorithms. 
For denoising of the profiled CIR we propose to use a bidirectional long short term memory (LSTM) network with computing blocks organized as in Figure~\ref{fig:bilstm}.
The bidirectional LSTM process the sequence data from past to future, forward direction, and from future to past, backward direction. Hence, being able to better learn the profiling pattern and remove noise.  
In our previous work~\cite{21VilascGAN}, we applied a LSTM network to correct quantization errors. Here, the implementation characteristics are similar as we train the model in a supervised learning fashion, but the task is slightly different.
For denoising of frequency domain signals, we propose to use untrained neural networks (UNNs) which implementation and advantages where explored in our recent work~\cite{22VilasUNN}.

\section{Model Order Selection} \
\label{sec:model_order}
In this section, we combine two blocks from Figure~\ref{fig:AI_FW}, the unitary ESPRIT prepossessing steps of the high-order singular value decomposition (HOSVD)~\cite{08HaardtTensor} and the ML instance for model order selection. The multidimensionality of MIMO OFDM wireless channels is better explored when using tensor algebra. In our previous work~\cite{21VilasMOS}, we have shown that using the $d$-mode singular values, computed by the HOSVD of the channels in the frequency domain, as input to a NN can enhance the model order selection accuracy and reduce the ML computational complexity if compared to state of the art solutions. Our ML for model order selection is trained in a supervised learning fashion where the NN is performing a multi-label classification task. For implementation details and detailed results, please refer to~\cite{21VilasMOS}.

 \section{Inference of Multipath Component Starting Parameters} 
\label{sec:start_parameters}
One of the main challenges for accurate parameter estimation in~\cite{profiling} is the computational complexity of the method. In this block of our framework, we aim to alleviate this complexity by employing a deep learning neural network to perform multipath component parameter estimation, this is delay, amplitude and phase. However, it is difficult to find a NN architecture that would best solve this mapping task. Hence, we use the NN's output parameters as starting point for the iterative profiling method~\cite{profiling}. 


The input signal to our MPC parameter estimation network is the profiled CIR $\mathbf{\bar{h}}_\mathrm{a}(\mathbf{t}_{\mathrm{s}},t_{i})$ with its absolute and phase values concatenated as 
$[~ |\mathbf{\bar{h}}_\mathrm{a}(\mathbf{t}_{\mathrm{s}},t_{i})| ~\sqcup_1~ \angle{\mathbf{\bar{h}}_\mathrm{a}(\mathbf{t}_{\mathrm{s}},t_{i})}~]^T~\in~\mathbb{R}^{{W} \times 2}$,
where $W$ is the size of the observation window. Table~\ref{tab:net-struc} presents the deep learning architecture for solving the parameter estimation. The size of the output layer depends on the model order, for $L$ MPCs there are $3L$ neurons in the output layer. Our deep network is trained using supervised learning with the cost function defined as
\begin{equation}
    \mathcal{L} = |  \Theta(t_{i})  -  \hat{\Theta}(t_{i}) | + \beta \mathcal{L}_\mathrm{prof},
\end{equation} 
where 
$\beta$ is a weighting factor and the profiling loss $\mathcal{L}_\mathrm{prof}$ is
\begin{equation}
    \mathcal{L}_\mathrm{prof} = \frac{|| \mathbf{\bar{h}}_\mathrm{a}(\mathbf{t}_{\mathrm{s}},t_{i}) - \mathbf{\hat{\bar{h}}}(\mathbf{t}_{\mathrm{s}},t_{i})||^2}{||\mathbf{\bar{h}}_\mathrm{a}(\mathbf{t}_{\mathrm{s}},t_{i})||^2},
\end{equation}
which computes the normalized squared error for reconstructing the profiled CIR.

\begin{table}[bt!]
\centering
\caption{\small{Description of the NN for MPC parameter estimation. Layer names refers to TensorFlow implementation.}}
\label{tab:net-struc}
\resizebox{0.45\columnwidth}{!}{%
\begin{tabular}{|c|c|c|c|}
\hline
 Layer & $N_\mathrm{filter}$ & Filter size & Activation \\ \hline
 Conv1D & $12$ & $3$ & ReLU \\ \hline
 MaxPool1D & 2 & - & - \\ \hline
 Conv1D & $12$ & $3$ & ReLU \\ \hline
 MaxPool1D & 4 & - & - \\ \hline
 Flatten & - & - & - \\ \hline
 Dense & $50$ & - & ReLU \\ \hline 
 Dense & $50$ & - & ReLU \\ \hline 
 Dense & $3L$ & - & Linear \\ \hline 
 \end{tabular}}
\end{table}

\section{Channel Prediction with Combined ML-PHY Framework} 
\label{sec:channel_prediction}

In this section we present the results of the interplay of the blocks \textit{ML instance for MPCs starting parameters} and \textit{iterative parameter estimation}. 
For that purpose, we create a synthetic dataset with $15000$ NLOS wireless channels with the MPCs delays and phase drawn from uniform distributions $U(0.15,5)\mathrm{T}_s$ and $U(0,2\pi)$~rad respectively, and the amplitudes drawn from an exponential distribution. From the MPCs parameters, we compute the profiled CIRs and preprocess them as described in Section~\ref{sec:start_parameters} with an observation window $W=256$ taps. We use the bidirectional LSTM to remove the profiling measurement noise and the model order selection block to define the number of output units for the NN estimating the starting parameters.

\begin{figure}[tb!]
    \centering
    \includegraphics[width=0.7\columnwidth]{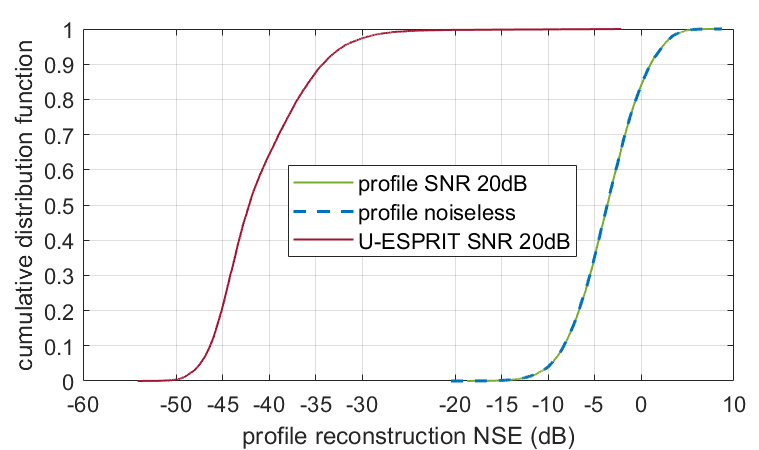}
    \caption{\small{Cumulative distribution function of the normalized squared error of the profiling reconstructed with the parameters $\hat{\Theta}(t_{i})$ estimated by our deep learning NN and by Unitary ESPRIT~\cite{uESPRIT}.}}
    \label{fig:profile-MLest}
\end{figure}

\begin{figure}
    \centering
    \includegraphics[width=\columnwidth]{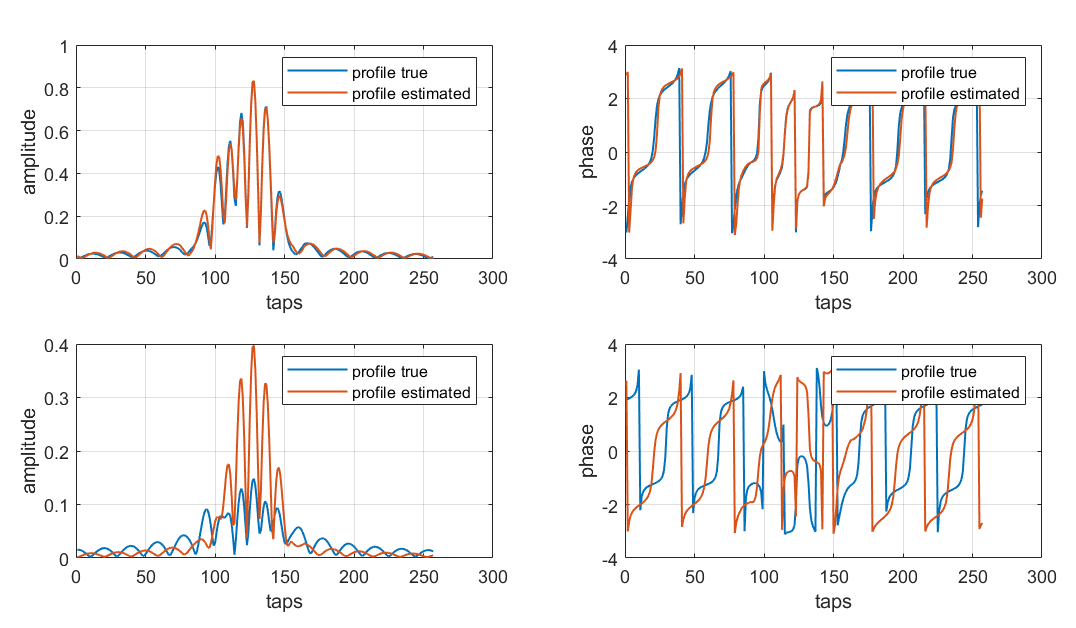}
    \caption{\small{Sample profiling signals reconstructed using $\hat{\Theta}(t_{i})$ from our NN. The first row presents the best case and the second row the worst case.}}
    \label{fig:profiling-sample}
\end{figure}

The deep learning network is trained using Python and TensorFlow. The learning rate of the Adam optimizer is set to $2 \times 10^{-3}$, and the cost function weighting factor is $\beta = 10$.   
We train our NN for profiled signals without noise, which we refer to as profile true or profile noiseless. Figure~\ref{fig:profile-MLest} presents the cumulative distribution function (CDF) of the normalized squared error (NSE) for the profile reconstructed with the starting parameters of our NN. 
The curve in blue is the error when considering a noiseless profiled signal as input to the NN, and the curve in green is when the profile input has a SNR of~$20$~dB, without retraining of the NN. 
We can observe that there is only a minor performance degradation of the ML method. 
Nonetheless, the profiling reconstruction error of our current NN is often high. Hence, there is a need for a further algorithm to refine the MPCs parameter estimates. In Figure~\ref{fig:profiling-sample}, we plot the comparison of two sample profiling, where our best NN profile reconstruction is in the first row, and our worst case is in the second row. 

\begin{figure}[htbp]
	\centering
	\includegraphics[width=0.7\columnwidth]{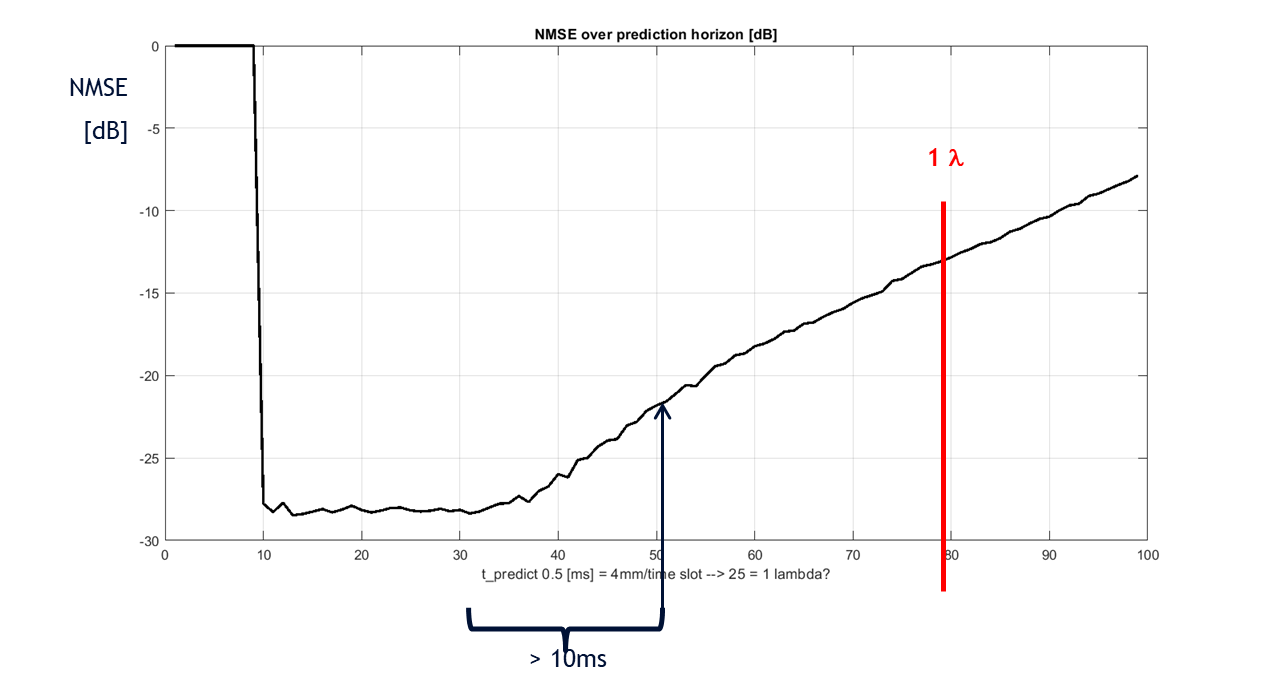}	
	\caption{\small{Channel prediction horizon for the proposed combined ML-PHY framework for the radio channel of the Nokia Campus mMIMO measurements at 2.136~GHz for a UE speed of 15~km/h.}}
	\label{fig:CP_results}
\end{figure}

\normalsize{
As mentioned in Section~\ref{sec:profiling}, for channel prediction we have to estimate accurate parameter sets $ \hat{\Theta}(t_{i}) $ for a number of time instances $ t_{i}~=~t_{1} \ldots t_{\mathrm{ob}} $, so that a proper extrapolation of the multipath component parameters is possible. For the evaluation of the best fitting concept, we have to consider three main issues, i.e., i) the achievable accuracy for the inference or estimation of the parameters, ii) the processing complexity, and iii) the overall latency. Note that, the last issue is especially relevant for channel prediction as the typical prediction horizon is in the range of 10~ms to maybe 100~ms. Hence, an algorithm with 
a latency significantly higher than 10~ms to 20~ms results in 
outdated and useless estimation parameters. 
Interestingly, the parameter estimation accuracy as such is less relevant for channel prediction, but instead we need a good matching of the profiled channel impulse responses, i.e., the loss function $ \mathcal{L}_\mathrm{prof} $ has to be small. The reason is, that for channel prediction we just have to find a smooth evolution of any parameter set  $ \hat{\Theta}(t_{i}) $, properly describing the evolution of the profiled channel impulse response. Note that 
this is different, for example, for the use case of positioning where we can accept a somewhat relaxed latency requirement, but require an extremely accurate parameter estimation, especially for the delay $ \tau_{l} $ of the strongest and/or LOS multipath component. 

The high complexity of the first test-wise implementations of the PHY based profiling method motivated the here proposed combined ML-PHY framework. In a first realization and for high accuracy, it took even hours for finding $ \hat{\Theta}(t_{i}) $ for a given profiled channel impulse response. In the meantime, a significant complexity reduction could be achieved by i) calculating in the time domain instead of
the
frequency domain corresponding to Equation~(\ref{equ:CIR_convolution_discrete}), ii) storing the often needed sinc-function into a memory and taking care of the relative delays $ \tau_{i} $ by a shifted readout of the stored sinc-function, iii) applying an extended search over the full parameter space only for the first profiled channel impulse response $ \mathbf{\bar{h}}^{'}(t_{i})|_{i=1} $
and, otherwise, apply a relative tracking to the so far found parameters over a limited search space. In addition, we search per multipath component $ l $ in each iteration $ j $ over the full set of all possible parameter variations $ \varDelta_{\kappa}~=~\{ -\Delta\tau \; 0 \; \Delta\tau, \ -\Delta\alpha \; 0 \; \Delta\alpha, \ -\Delta\varphi \; 0 \; \Delta\varphi  \} $, which results in overall $ 3 \times 3~=~27 $ parameter variations. The index $ \kappa $ defines different values for $ \Delta\tau $, $ \Delta\alpha $, and $ \Delta\varphi  $
so that the search space is either \emph{coarse}, \emph{medium}, or, \emph{fine}:}
\begin{equation}
\small{
	\begin{array}{rrl} 
	\label{equ:parameters_es}
		\Theta^{j,\kappa}_{l}(t_{i}) & = & \{ \tau_{1} \alpha_{1} \varphi_{1}, \ldots,   \tau_{l} \alpha_{l} \varphi_{l} + \varDelta_{\kappa},  \ldots, \tau_{L} \alpha_{L} \varphi_{L}  \}  |_{t = t_{i}} ;
	\end{array}}
	\end{equation}
\normalsize{This time domain implementation is very fast as it includes only multiplications and sum operations.} To evaluate the complexity of our above proposed methods, we implemented all algorithms and all NNs in MatLab and used the 'tic' and 'toc' function on a conventional computer with a Intel core i9 main board. Note that nowadays counting of floating point operations (FLOPS) is less common due to the more complex compute architectures and methods. 
The results are given in Table~\ref{tab:processing_time}, where \emph{profiling start} is for the fully iterative estimation of $ \mathbf{\bar{h}}^{'}(t_{i})|_{i=1} $,  \emph{profiling tracking} is for the tracking over ten time steps a 2~ms of $ \mathbf{\bar{h}}^{'}(t_{i})|_{i=2 \ldots 10} $ relative to the first estimation, \emph{NN start parameter inference} is the NN for inference of the start parameters, and \emph{Unitary ESPRIT} for the direct estimation of $ \mathbf{\bar{h}}^{'}(t_{i}) $ for any time instant $ i $. The table provides the required delays as well as the achieved accuracy levels. 
\begin{table}[bt!]
	\centering
	\caption{\small{Processing time and NMSE for proposed methods.}}
	\label{tab:processing_time}
	\resizebox{0.9\columnwidth}{!}{%
		\begin{tabular}{|c|c|c|}
			\hline 
			\emph{profiling start} & $ 80~ms $ &   -40~dB  \\ \hline
	    	\emph{profiling tracking} & $ 15~ms $ & -35~dB  \\ \hline
	    	\emph{NN start parameter inference} & $ 4~ms $ & avg -5~dB (see CDF)  \\ \hline
	    	\emph{Unitary ESPRIT}~\cite{uESPRIT} & $ 106~ms $ & avg -40~dB   \\ \hline
		\end{tabular}}
	\end{table}
From the given results we identify a first useful setup for our proposed AI/ML framework. As \emph{profiling start} has a high latency of $80$~ms, we propose to infer first a set of start parameters by \emph{NN start parameter inference}. The accuracy is then just around $-5$~dB so that we add at least one \emph{profiling tracking} step leading to the overall delay of $4$~ms $+~ 15$~ms~$=~19$~ms. This is at least close to our target of $10$~ms. When the first parameter set $ \mathbf{\bar{h}}^{'}(t_{i})|_{i=1} $ is found, we switch into tracking mode with about $15$~ms estimation time per time step, while the NMSE is still around $-35$~dB. Unitary ESPRIT would provide an accurate first parameter set $ \mathbf{\bar{h}}^{'}(t_{i})|_{i=1} $, but 
is less useful due to the large processing time of $106$~ms due to the SVD computation and matrix inversion. 
Figure~\ref{fig:CP_results} illustrates the achieved channel prediction performance, where the channel, as described in Section~ \ref{sec:framework}, is observed at time instances 
$ t_{i}~=~10$ to $30$ to do a channel prediction from 
$ t_{i} ~=~31$ to $100$.

\section{Conclusion} \
\label{sec:conclusion}
We have proposed 
a
channel prediction framework consisting of a combination of suitably chosen AI/ML and rule based PHY layer algorithms. 
Currently,
our results indicate that neural networks like bidirectional LSTMs or UNNs are well suited for the purpose of noise reduction. Promising is the combination of a first ML based estimate of multipath component start parameters with the iterative PHY based fine tuning and tracking of these parameters. With about $15$~ms to $20$~ms the latency is close to the intended 
$10$~ms. Next, we will consider further refinements of the algorithms and parallel processing including GPUs.

\footnotesize{
\section*{Acknowledgment}
This research was partly funded by German Ministry of Education and Research (BMBF) under grant 16KIS1184 (FunKI).
The measured massive MIMO radio channels have been provided by our Nokia Bell Labs colleagues from Stuttgart. Many thanks to Stefan Wesemann for his direct support.  }

\bibliographystyle{IEEEtran}
\bibliography{REF_SPAWC_4_1}
\end{document}